\documentclass[11pt,twoside]{article}
\usepackage{./asp2014}

\aspSuppressVolSlug
\resetcounters

\begin{document}

\title{On the binary origin of FS CMa stars: young massive clusters as test beds}
\author{Diego de la Fuente,$^1$ Francisco Najarro,$^1$ and Miriam Garcia$^1$
\affil{$^1$ Centro de Astrobiolog\'ia (CSIC/INTA), ctra. de Ajalvir km. 4, 28850 Torrej\'on de Ardoz, Madrid, Spain; \email{delafuente@cab.inta-csic.es}}}

\paperauthor{Diego de la Fuente}{delafuente@cab.inta-csic.es}{}{Centro de Astrobiolog\'ia}{Departamento de Astrof\'isica}{Torrej\'on de Ardoz}{Madrid}{28850}{Spain}
\paperauthor{Francisco Najarro}{najarro@cab.inta-csic.es}{}{Centro de Astrobiolog\'ia}{Departamento de Astrof\'isica}{Torrej\'on de Ardoz}{Madrid}{28850}{Spain}
\paperauthor{Miriam Garcia}{mgg@cab.inta-csic.es}{}{Centro de Astrobiolog\'ia}{Departamento de Astrof\'isica}{Torrej\'on de Ardoz}{Madrid}{28850}{Spain}

\begin{abstract}
FS CMa stars are low-luminosity objects showing the B[e] phenomenon whose evolutionary origin is yet to be unraveled. Various binary-related hypotheses have been recently proposed, two of them involving the spiral-in evolution of the binary orbit. The latter occurs more often in dense stellar environments like young massive clusters (YMCs). Hence, a systematic study of FS CMa stars in YMCs would be crucial to find out how these objects are created. In YMCs, two FS CMa stars have been confirmed and three candidates have been found through a search method based on narrow-band photometry at Paschen-$\alpha$ and the neighboring continuum. We apply this method to archival data from the Paschen-$\alpha$ survey of the Galactic Center region, yielding a new candidate in the Quintuplet cluster. Limitations of this method and other alternatives are briefly discussed.
\end{abstract}

\section{Introduction}
  \label{sec:intro}

FS CMa stars are defined by \citet{miroshnichenko07} as a subtype of objects showing the B[e] phenomenon that fulfills the following observational features: B or early-A spectral types; low luminosity ($2.5 \lesssim log(L/L_\odot) \lesssim 4.5$); infrared excess steeply decreasing at $\lambda \sim 20~\mu \mathrm{m}$; and location outside sites of ongoing star formation. These features suggest that FS CMa stars may be main-sequence-like objects surrounded by compact disks of warm dust whose origin is not related with star formation.

FS CMa stars are very puzzling objects, regarding both the evolutionary state and the ejection mechanism of the circumstellar matter. Observationally derived mass-loss rates \citep{miroshnichenko+00, carciofi+10} would be $\sim 2$~dex higher than predicted by the stellar wind theory if we assumed they are single stars.

Recently, \citet{delafuente+15} have presented the first confirmation of coeval stellar populations hosting FS CMa stars. Specifically, two FS CMa stars (Mc20-16 and Mc70-14) were serendipitously detected in two young massive clusters (YMCs), Mercer 20 and Mercer 70, thanks to two apparently contradicting photometric features on NICMOS/HST images. First, they were higlighted as strong emitters in the Paschen-$\alpha$ line ($P_\alpha$, $\lambda=1.875 \mu~\mathrm{m}$), as measured by subtraction of images of two narrow-band filters, F187 and F190N (located at the $P_\alpha$ wavelength and the neighboring continuum, respectively). Second, these objects appeared unexpectedly faint on the F222M broad-band image. After spectroscopic confirmation of the two newly-discovered objects, cluster membership and coevality allowed the first age determination for the two FS CMa stars, being incompatible with the hypothesis of binary post-AGB stars (see Sect. \ref{sect:channels}).

\section{Formation channels and the role of dense clusters}
  \label{sect:channels}

As mentioned in Sect. \ref{sec:intro}, any hypothesis for the nature of FS CMa stars based on the evolution of single stars is seriously challenged by the apparently high mass-loss rates. Therefore, it is hard to explain how the B[e] phenomenon arise from single main-sequence stars. Conversely, several binary-related scenarios for the origin of these objets have been recently proposed:

\begin{enumerate}
 \item A close binary system that has recently finished a rapid phase of mass transfer, leaving a circumstellar or circumbinary dusty disk \citep{miroshnichenko07}.
 
 \item An intermediate-mass post-AGB star in a binary system, in such a way that the planetary nebula ejection is delayed due to dynamical interaction with the secondary star \citep{miroshnichenko+13}.
 
 \item A binary merger product surrounded by a disk of matter that was expelled during the merging process \citep{delafuente+15}.
\end{enumerate}

In principle, these scenarios do not have to be mutually exclusive. It is still possible that the FS CMa class comprises objects of similar observational features but were created through different evolutionary channels. Having this in mind, the post-AGB binary nature (\#2) can only be discarded for Mc20-16 and Mc70-14, being still possible for other FS CMa stars. On the other hand, the rest of the above listed hypotheses (\#1 and \#3) can be induced by the evolution of the binary orbit, especially when both components are still in the main sequence (which seems likely for FS CMa stars). Hence, these two scenarios are favored in dense clusters, where orbit spiraling-in may be enhanced due to dynamical interactions with other cluster members.

These considerations give us some clues about the strategy to adopt for elucidating the real origin of FS CMa stars. First, if these objects proceed from evolution of the binary orbit (scenario \#1 or \#3), then FS CMa stars will be more common in clusters than in the field. Second, we can distinguish between the recent mass transfer (\#1) and the merger (\#3) hypothesis by examining the fraction of close binaries among FS CMa stars: the merger channel would yield a much lower close binary fraction\footnote{Merger products are not necessarily single: triple stars can also be merger progenitors.}. Finally, any systematic dependence of stellar/disk physical properties on the environment would indicate that different formation channels could be sharing the responsibility of creating FS CMa stars.

\section{Searching for new FS CMa candidates}

The discussion in Sect. \ref{sect:channels} makes clear the benefits of demographic studies of FS CMa stars in clusters. Although only two cases have been confirmed in clusters so far, \citet{delafuente+15} propose a method for finding new candidates in a particular type of YMCs, specifically those that host Wolf-Rayet (WR) stars. This method is based on the strong line emission of FS CMa stars, comparable to WR stars, while having a much lower luminosity. Provided that they belong to the same coeval population, it is possible to define a region in the $P_\alpha$-magnitude diagram of the cluster where FS CMa candidates are pinpointed \citep[see][Fig. 13]{delafuente+15}.

\articlefigurefour{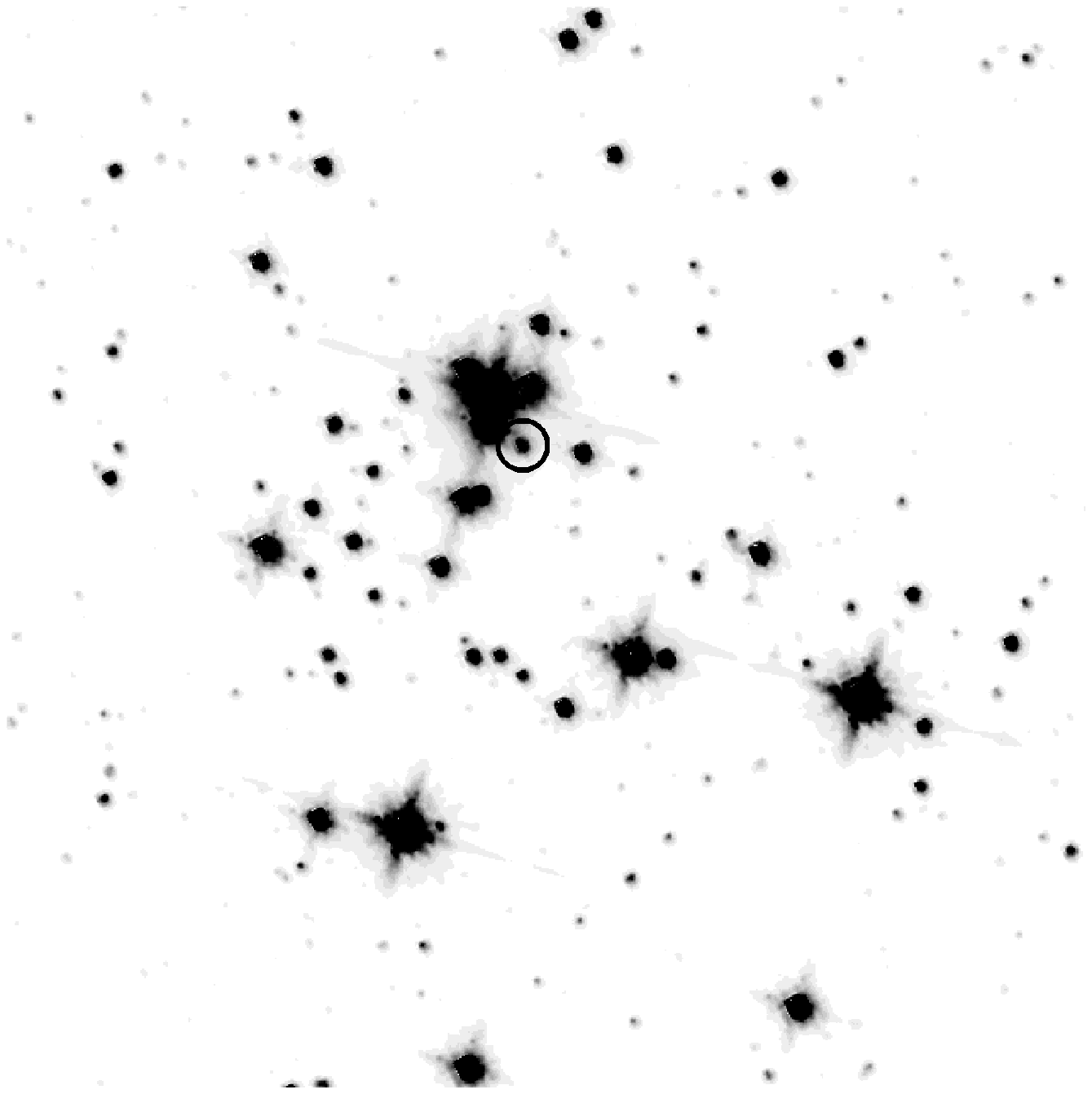}{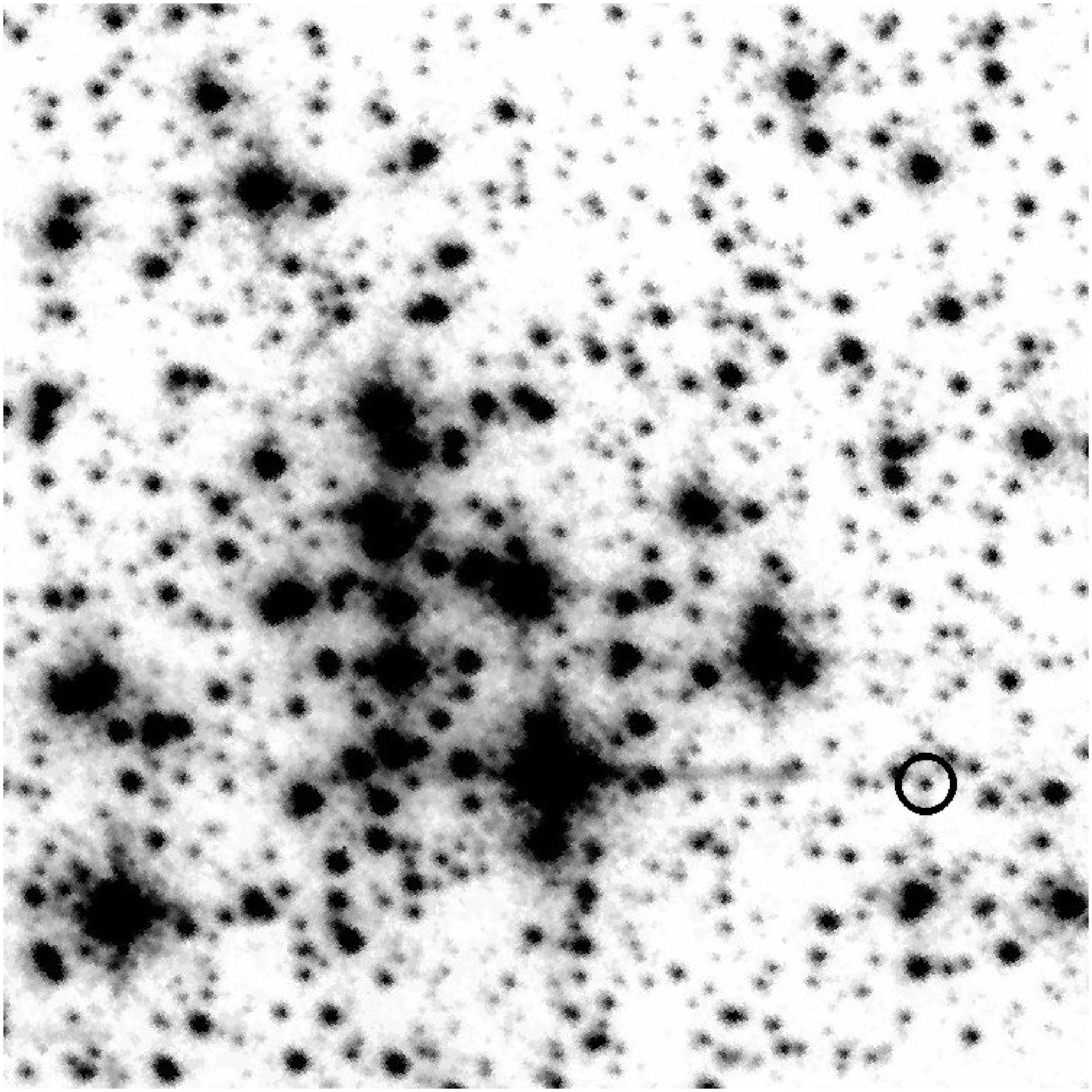}{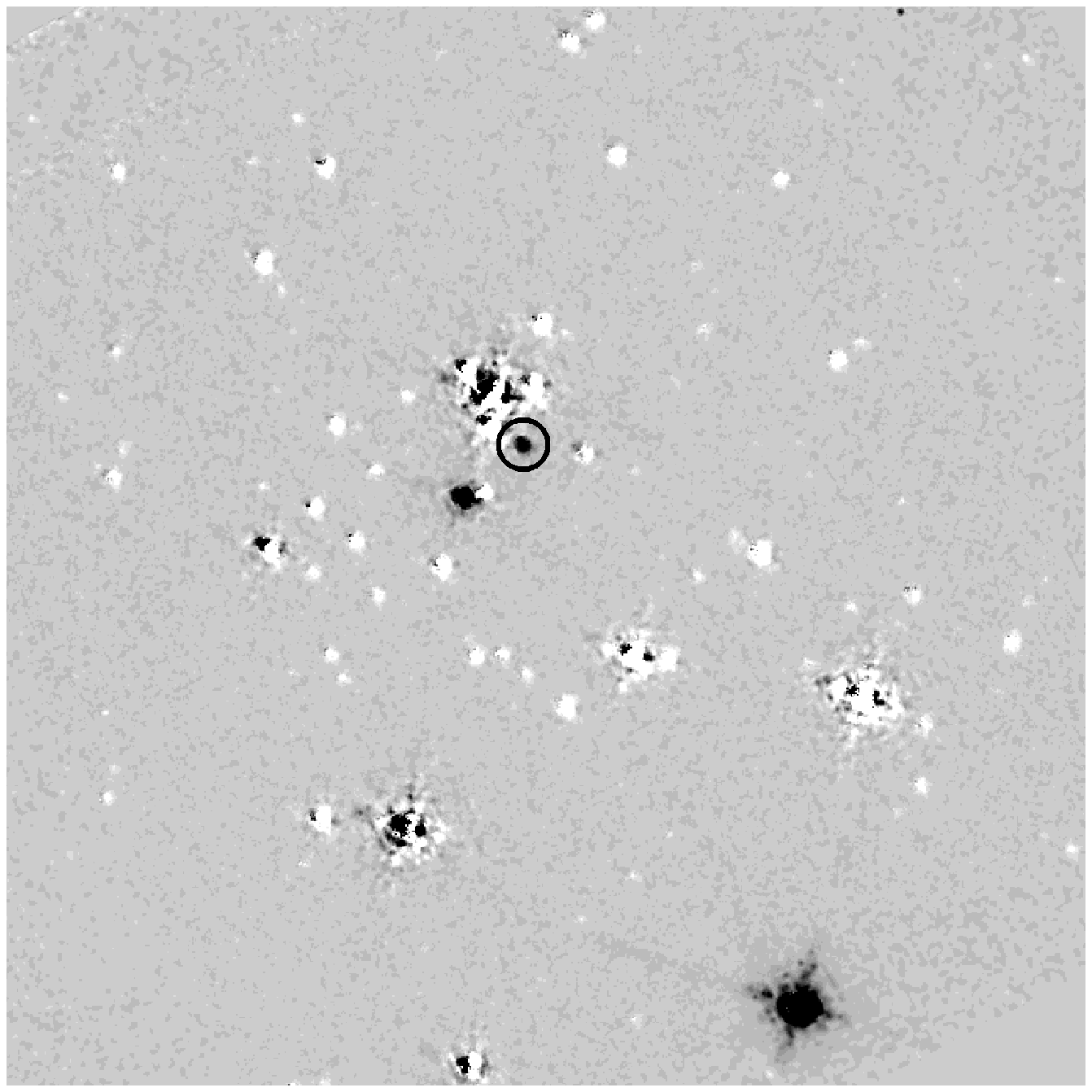}{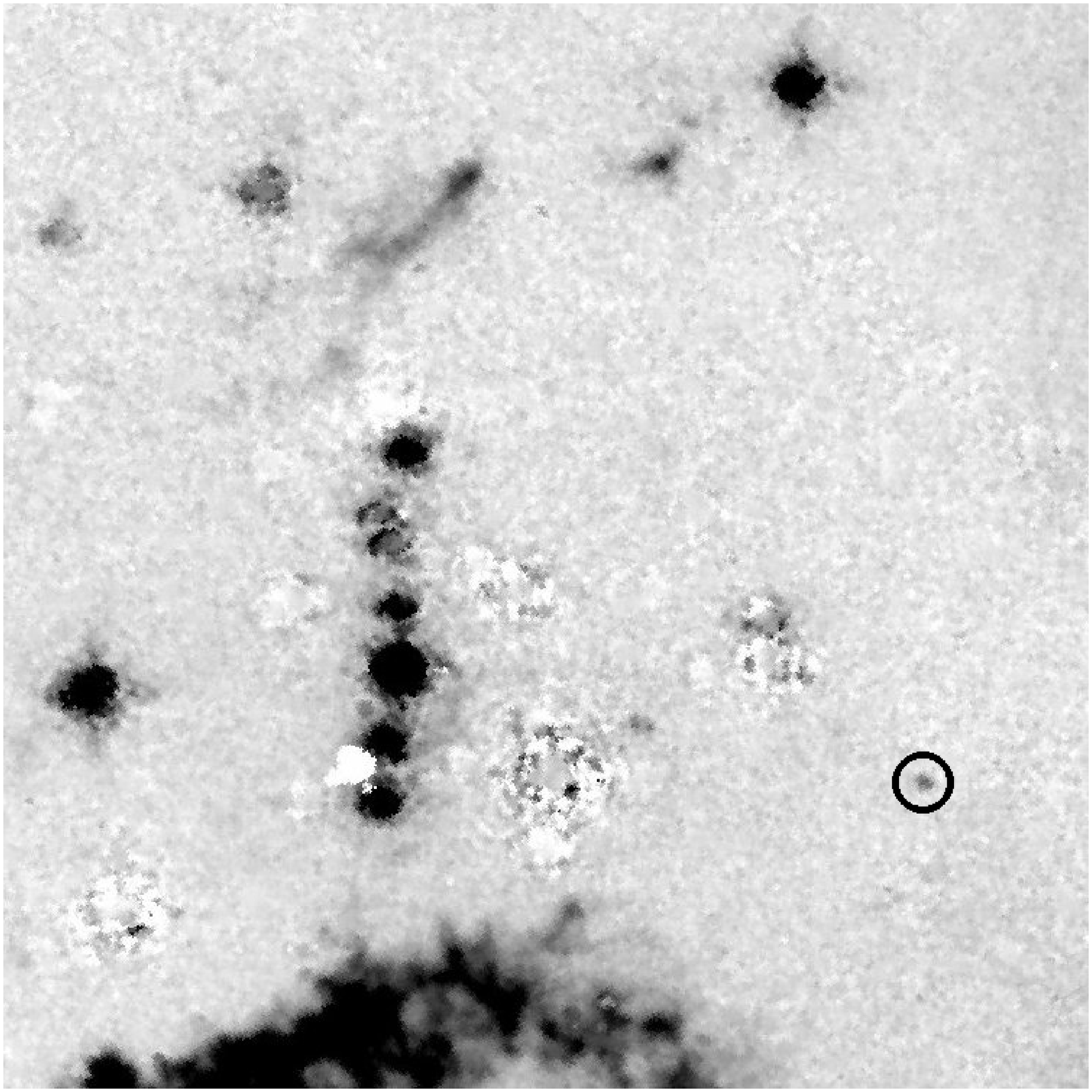}{fig:quintuplet}{NICMOS/HST images of Mercer 70 (\textit{left pannels}) and the Quintuplet cluster (\textit{right pannels}); their sizes are $1.58 \times 1.58$~pc assuming distances of 7.0 kpc for Mercer 70 \citep{delafuente+15} and 8.5 kpc for the Galactic Center region \citep{kerr-lyndenbell86}. The \textit{upper pannels} display single images in the F222M and F190N bands respectively, while the \textit{lower pannels} show the F187N$-$F190N subtraction (i.e. the $P_\alpha$ images). The confirmed FS CMa star in Mercer 70 and the new candidate in the Quintuplet cluster are circled. North is up and East is left.}

\citet{delafuente+15} only applied the aforementioned method to clusters that were observed as part of the same HST program than Mercer 20 and Mercer 70, yielding three new FS CMa candidates in Danks 1 and Mercer 81. In this contribution, we apply the search method to the Paschen-$\alpha$ survey of the Galactic Center region \citep{wang+10, dong+11}, whose spatial coverage includes two coeval YMCs hosting WR stars, namely the Quintuplet and the Arches clusters. Stellar sources in the \citet{dong+11} photometric catalog are classified in three groups, according to the $F187N-F190N$ measurement: primary $P_\alpha$ emitters, with $S/N > 4.5$; secondary $P_\alpha$ emitters, whose emitter status is less reliable due to low $S/N$ or other problems (e.g. contamination from a neighboring source); and non-emitters.

\articlefigure{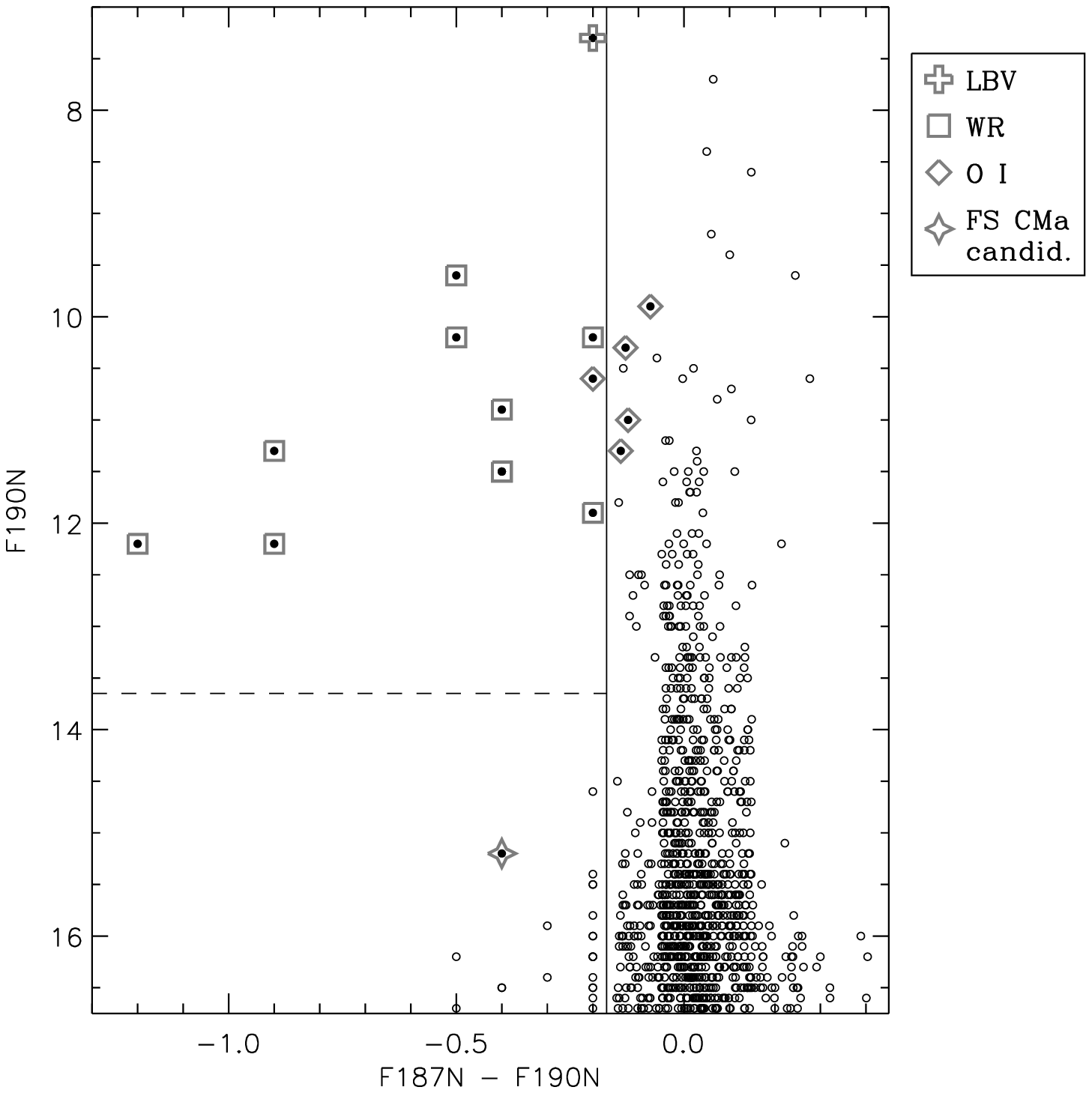}{fig:cmd}{$P_\alpha$-magnitude diagram of the Quintuplet cluster ($1'$ diameter field centered at $l=0.165$, $b=-0.06$). Filled dots are considered as primary $P_\alpha$ emitters by \citet{dong+11}, while open dots are other stellar sources. Since \citet{dong+11} only provides magnitudes to the first decimal place, many data points would be coincident on crowded parts of the diagram; in order to avoid underrepresentation, a random abscissa value between $-0.05$ and $+0.05$ is added to each data point fulfilling $F187N-F190N > -0.2$. The solid line shows the $P_\alpha$-excess limit as adopted by \citet{delafuente+15}, and the dashed line separates the WR and FS CMa luminosity regimes.}

When examining the \citep{wang+10} subtracted images of the Quintuplet cluster, we find an object that is clearly visible in the F187N$-$F190N subtraction whose F190N counterpart is much fainter than the remaining line-emitters, thus being consistent with a FS CMa candidate designation (see comparison with the Mercer 70 images in Figure \ref{fig:quintuplet}). It is listed as the primary $P_\alpha$ emitter \#74 at \citet[Table 3]{dong+11}. Additionally, we build the $P_\alpha$-magnitude diagram of the Quintuplet cluster (Figure \ref{fig:cmd}), where all primary $P_\alpha$ emitters are highlighted with the spectral types compiled by \citet{dong+11}.
Object \#74 turns out to be the only primary $P_\alpha$ emitter in the region of the diagram where only FS CMa stars are expected ($F187-F190N < -0.17; F190N > 13.65$, the latter being approximately equivalent to $M_K > -4$, assuming an extinction of $A_\mathrm{F190N}=3.0$ and a distance of 8.5 kpc). Although a few more sources are present in this region of the diagram, they are listed by \citet{dong+11} as non-emitting sources (i.e. with no reliable $P_\alpha$ emission), therefore they are discarded as FS CMa candidates. 

On the other hand, we have found no FS CMa candidate in the Arches cluster. Nevertheless, it is important to be aware of the extremely crowded appearance of the Arches Cluster even on the full-resolution NICMOS/HST images, which would mask any faint object in the central region due to source confusion.

\begin{table}[!t]
\caption{Coeval clusters hosting WR stars with available Paschen-$\alpha$ photometry.}
  \label{tab:ymcs}
\smallskip
\begin{center}
{\small
\begin{tabular}{lcccccc}  
\tableline
\noalign{\smallskip}
Cluster ID & $l$ & $b$ & Age [Myr] & Mass [$M_\odot$]& References & $N_{c}$~$^a$ \\
\noalign{\smallskip}
\tableline
\noalign{\smallskip}
Arches     & 0.123   & $+0.018$ & $2.5 \pm 0.5$       & $(4.9 \pm 0.8) \times 10^4$ & 6, 9, 11 & 0  \\
Danks 1    & 305.338 & $+0.072$ & $1.5^{+1.5}_{-0.5}$ & $(8.0 \pm 1.5) \times 10^3$ & 1        & 2  \\
Danks 2    & 305.393 & $+0.087$ & $3^{+3}_{-1}$       & $(3.0 \pm 0.8) \times 10^3$ & 1        & 0   \\
Mercer 20~$^b$ & 44.170  & $-0.069$ & $[5.0,6.5]$     & $>2.0 \times 10^4 $         & 3 & \textbf{1} \\
Mercer 23  & 53.770  & $+0.164$ & $4.0^{+1.7}_{-0.3}$ & $3-5 \times 10^3$           & 5, 8     & 0    \\
Mercer 30  & 298.755 & $-0.408$ & $4.0 \pm 0.8$       & $(1.6 \pm 0.6) \times 10^4$ & 4        & 0   \\
Mercer 70~$^b$ & 329.697 & $+0.583$ & $4.5 \pm 1.0$   & $> 1.4 \times 10^4$         & 3 & \textbf{1} \\ 
Mercer 81  & 338.395 & $+0.102$ & $3.7^{+0.7}_{-0.5}$ & $\sim 3 \times 10^4$        & 2, 3     & 1   \\
Quintuplet & 0.165   & $-0.061$ & $3.5 \pm 0.5$       & $1.6 \times 10^4$           & 6, 7, 10 & 1    \\
\noalign{\smallskip}
\tableline
\end{tabular}
}
\end{center}
\small \textbf{References.} 1, \citet{davies+12a}; 2, \citet{davies+12b}; 3, \citet{delafuente+15};  4, \citet{delafuente+16}; 5, de la Fuente et al. (in prep.); 6, \citet{dong+11}; 7, \citet{figer+99}; 8, \citet{hanson+10}; 9, \citet{harfst+10}; 10, \citet{liermann+12}; 11, \citet{najarro+04}.\\
$(^a)$ Number of FS CMa candidates; those confirmed by spectroscopy are shown in bold.\\
$(^b)$ Ages and masses of Mercer 20 and Mercer 70 are slightly reviewed, relative to values in \citet{delafuente+15}, thanks to improved isochrone fitting.
\par
\end{table}

We present in Table \ref{tab:ymcs} an updated list of YMCs where a photometric search for FS CMa candidates has been carried out. We note that spectroscopic confirmation is still pending for the non-confirmed candidates. From these data we estimate that the detection rate of candidate FS CMa stars in YMCs with 1.5-7 Myr old (when WR stars are expected) is $\sim 0.4$ per $10^4~M_\odot$. However this is a lower limit, since we are only detecting FS CMa candidates on the upper part of the expected luminosity range. Faint line-emitters are strongly affected by uncompleteness, either because of photometric uncertainties (e.g. due to insufficient S/N in F187N$-$190N, as at the bottom of Fig. \ref{fig:cmd}), or due to stellar crowding.

\section{Conclusions and future work}

We update the census of FS CMa candidates in YMCs with a new object in the Quintuplet cluster. Also, we argue that demographics of FS CMa stars in dense clusters relative to the field can provide crucial hints on the binary origin of FS CMa stars. $P_\alpha$ photometry of YMCs has been proven very useful for this purpose, despite incompleteness effects.

On the one hand, spectroscopic follow-up is needed for the FS CMa candidates in Table \ref{tab:ymcs}, in order to validate the \citet{delafuente+15} photometric method. On the other hand, further $P_\alpha$ observations of YMCs should be carried out, but unfortunately, the only instrument having filters in this line and the adjacent continuum, NICMOS/HST, is out of operation. Until the advent of new suitable instrumentation (e.g. NIRCam/JWST), we are studying the use of other near-infrared narrow-band filters. However, these alternatives might be less convenient for the required deep photometry. For example, narrow-band imaging at the Paschen-$\beta$ line can be suitable only for relatively close, not too extincted YMCs, as this line is considerably more afected by reddening than $P_\alpha$; we could also use Brackett-$\gamma$, although this line is expected to be significantly weaker than Paschen-$\beta$.

\acknowledgements This research has been supported by the Spanish government through project ESP2013-47809-C3-1-R.

\bibliography{biblio}

\end{document}